\documentclass{ws-mpla}
\usepackage[super,compress]{cite}
\usepackage{amsmath}
\usepackage{indentfirst}
\usepackage{graphicx}
\usepackage[all]{xy}
\usepackage{float}

\begin{document}

\renewcommand{\draftnote}{}
\renewcommand{\trimmarks}{}

\markboth{I. Alikhanov}{Probing the Glashow resonance at electron--positron colliders}

%%%%%%%%%%%%%%%%%%%%% Publisher's Area please ignore %%%%%%%%%%%%%%%
%
\catchline{}{}{}{}{}
%
%%%%%%%%%%%%%%%%%%%%%%%%%%%%%%%%%%%%%%%%%%%%%%%%%%%%%%%%%%%%%%%%%%%%

\title{Probing the Glashow resonance at electron--positron colliders}

\author{I. Alikhanov}

\address{Institute for Nuclear Research of the Russian Academy of Sciences,\\ Moscow, 117312, Russia\\
Institute of Applied Mathematics and Automation KBSC RAS,\\ Nalchik, 360000, Russia\\ialspbu@gmail.com}

\maketitle
\pub{}{}
%\begin{history}
%\received{Day Month Year}
%\revised{Day Month Year}
%\end{history}

\begin{abstract}
The Glashow resonance is a rapid enhancement of the cross section for scattering of electron antineutrinos on electrons at the $W^-$ boson production threshold due to the $s$-channel contribution. This resonance is being searched for at large volume neutrino detectors in which the reaction $\bar\nu_ee^-\rightarrow W^-$ to be initiated by cosmic ray antineutrinos of energies about 6.3~PeV. The relatively small neutrino flux reaching the Earth together with the necessity of analyzing ultra-high-energy final states make such searches challenging. 
We argue here that the Glashow resonance may contribute to the process~$e^+e^-\rightarrow~W^+W^-$. By extending the method of the effective (equivalent) particles to the neutral leptons, we relate the distribution of the  effective neutrinos in the electron to the total cross section for~$e^+e^-\rightarrow~W^+W^-$. Our approach gives a good fit to existing experimental data measured at the collider LEP at CERN and allows one to interpret these measurements as an observation of the Glashow resonance. We also discuss an advantage of future electron--positron colliders for probing the resonance.

\end{abstract}

\ccode{PACS numbers: 13.15.+g, 14.70.Fm, 14.60.-z}

%\tableofcontents

%%%%%%%%%%%%%%%%%%%%%%
\section{Introduction}
%%%%%%%%%%%%%%%%%%%%%%
Neutrinos are likely the most poorly understood basic constituents of the Standard Model. It is a challenging problem for experiments to create high intensity and well-collimated neutrino beams with known flavor compositions in order to probe neutrino properties. For this reason there is a lack of convincing  evidences for numerous important processes of neutrino interactions at high energies predicted by the theory. This is the case for $\bar\nu_ee^-$~scattering near the $W^-$~boson production threshold, where a resonant enhancement of the corresponding cross section is expected. This effect was predicted by Glashow in 1959 and commonly referred to as the Glashow resonance~\cite{Glashow_res}.

The strategy of searching for this resonance adopted today is using large volume water/ice neutrino detectors~\cite{berezinsky} in which the reaction $\bar\nu_ee^-\rightarrow W^-$ to be initiated by cosmic ray antineutrinos of energies about $m_W^2/(2m_e)=6.3~\text{PeV}$ ($1~\text{PeV}=10^{15}~\text{eV}$). This channel is of particular interest for astrophysics because can shed light on the~$\bar\nu_e$~content of cosmic rays at PeV energies and thereby on mechanisms of the neutrino production at the source and propagation to the Earth~\cite{Glashow_res_cube1, Glashow_res_cube2, Glashow_res_cube3, Glashow_res_cube5,Glashow_res_cube6,Biehl:2016psj}. The  IceCube neutrino telescope~\cite{icecube_rev} has already detected four neutrino events in the PeV region~\cite{icecube_2events,icecube_28,icecube_new,Aartsen:2016xlq} that might be from the Glashow resonance~\cite{Glashow_res_cube7,Glashow_res_cube8}. Three of the events are showers initiated and contained in the detector with reconstructed energies in the interval 1--2~PeV and one is a track went through the detector depositing $2.6\pm0.3$~PeV therein. However, the statistical significance of these data is not still sufficient to make a decisive conclusion~\cite{Barger:2014iua,Kistler:2016ask,Sahu:2016qet,Gauld:2019pgt,Huang:2019hgs}. Apart from the statistical reason, the identification of  the on-shell $W$ events in large volume neutrino detectors is also subject to the assumptions about the form of the neutrino spectrum, the lepton flavor content of the neutrino flux, the types of the particles giving the through-going tracks~\cite{Kistler:2016ask} and possible additional channels of $W$ boson production in neutrino--nucleus collisions\cite{Seckel:1997kk,Alikhanov:2014uja,Alikhanov:2015kla,Beacom:2019pzs}. With the commissioning of the new detectors KM3NeT~\cite{Adrian-Martinez:2016fdl}, Baikal-GVD~\cite{Avrorin:2018ijk} and  IceCube-Gen2~\cite{Aartsen:2014njl}, the problem of observation of the Glashow resonance excited by the cosmic neutrinos will hopefully be resolved. 

Meanwhile, there are processes in which a resonance is excited without manifesting itself by a peak in the cross section~\cite{Alikhanov:2013fda}. This occurs when one of the colliding particles ($\bar\nu_e$~or~$e^-$) does not have a definite momentum but characterized instead by a distribution of the momenta, like partons in the nucleon. For example, the $W$ bosons can be resonantly produced in this a way by neutrinos scattering on photons or atomic nuclei~\cite{Alikhanov:2014uja,Alikhanov:2015kla,Alikhanov:2008eu}. At the same time the required neutrino energies may be far below the PeV region. Furthermore, such reactions can give access to measurements of the resonant scattering of all the three neutrino flavors $\nu_e$, $\nu_\mu$ and $\nu_\tau$~\cite{Alikhanov:2015kla}.

Recently it was proposed to extend the concept of the effective (equivalent) particle~\cite{Chen:1975sh} to neutrinos~\cite{Alikhanov:2018xdi}. In analogy with the quark densities in the quark--parton model~\cite{Bjorken:1969ja}, one can introduce distributions of the effective neutrinos in charged leptons. The effective neutrino approximation~(ENA) may provide a framework for probing neutrino-induced reactions at electron--positron colliders as well as at other lepton colliding facilities. This is similar to studies of quark--quark interactions in hadron--hadron collisions through Drell--Yan processes~\cite{Drell:1970wh} or of two photon physics at $e^+e^-$ colliders relying on the Weizs\"acker--Williams approximation~\cite{Brodsky:1971ud,Walsh:1973mz,Budnev:1974de}.
For example, it was shown that in the high energy limit one may envisage the process $e^+e^-\rightarrow W^+W^-$ as proceeding through the Glashow resonance~\cite{Alikhanov:2018xdi}. In this article we investigate a possible contribution from the resonance in the whole range of center-of-mass (cms) energies starting from the threshold~$\sqrt{s}=2m_W$.  

The article is organized as follows. In Sec.~\ref{eweak} we remind the structure of the total cross section for~$e^+e^-\rightarrow~W^+W^-$ in the electroweak theory. In Sec.~\ref{ena} we calculate the cross section within the ENA and compare this with the exact result. In Sec.~\ref{experiment} the prediction of the effective neutrino approach is compared with existing experimental data. Sec.~\ref{concl} contains a summary and the conclusions.

%%%%%%%%%%%%%%
% FIGURE 1
%%%%%%%%%%%%%%
\begin{figure}[t]
\centering
\includegraphics[width=.8\textwidth]{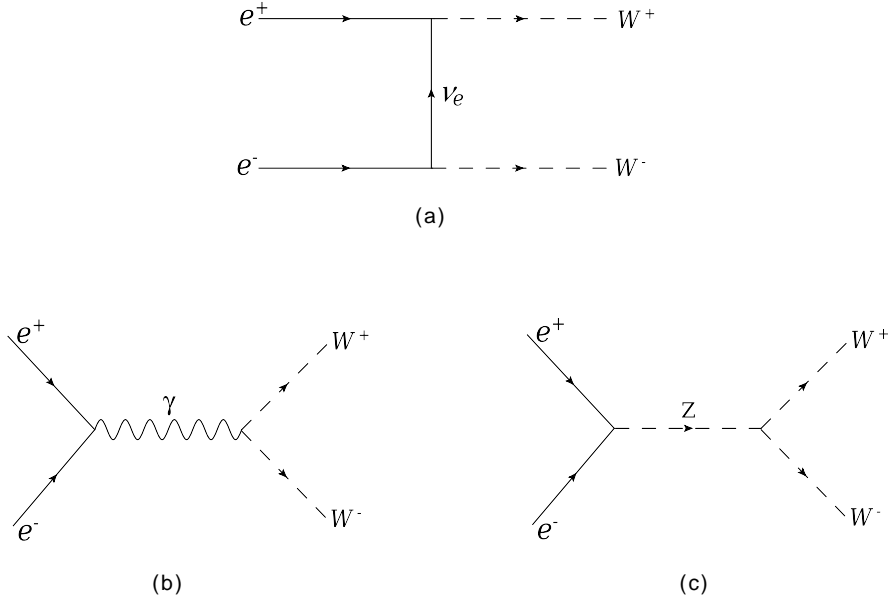}
\caption{The lowest order Feynman diagrams that contribute to $e^+e^-\rightarrow W^+W^-$ in the massless electron limit.}
\label{leading}
\end{figure}
%%%%%%%%%%%%%
%%%%%%%%%%%%%%%%%
\section{The process $e^+e^-\rightarrow W^+W^-$ in the electroweak theory \label{eweak}}
\begin{figure}[t]
\centering
\includegraphics[width=0.7\textwidth]{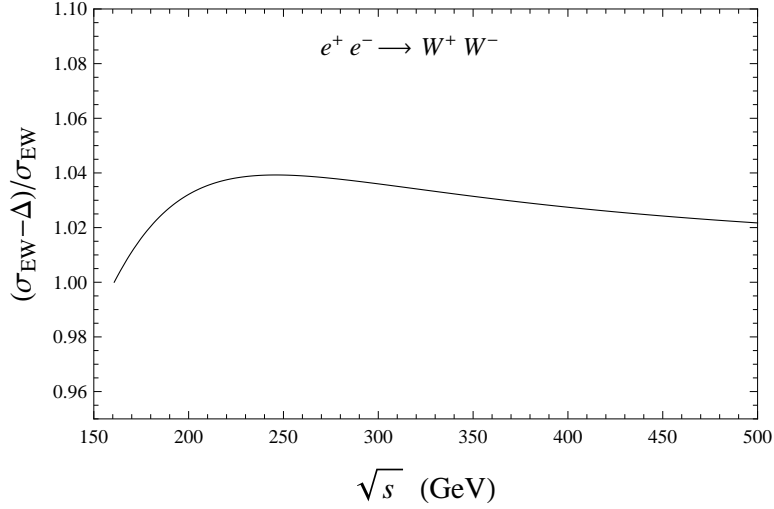}
\caption{This plot shows that the relative contribution of the term $\Delta$ to the total cross section for $e^+e^-\rightarrow W^-W^+$, Eq.~\eqref{funfel}, is~$\lesssim4\%$. On the abscissa is the cms energy.}
\label{fig_ratio}
\end{figure}

%%%%%%%%%%%%%%%%%
The lowest order Feynman diagrams that contribute to~$e^+e^-\rightarrow W^+W^-$ in the massless electron limit are depicted in~Fig.~\ref{leading}. The corresponding total cross section  in the electroweak theory has been calculated long ago~\cite{Flambaum:1974wp,Alles:1976qv} and can be represented as

\begin{equation}
\sigma_{\text{EW}}=\frac{\pi\alpha^2}{2\sin^4{\theta_W}}\frac{1}{s}\left[\frac{1+\tau-2\tau^3}{1-\tau}\ln\left(\frac{1+\beta}{1-\beta}\right)-\frac{5}{4}\beta\right]+\Delta,
\label{funfel}
\end{equation}
where $\alpha$ is the fine structure constant, $\theta_W$ is the Weinberg angle, $\beta=\sqrt{1-4m_W^2/s}$, $\tau=~m^2_W/s$ with $m_W$ and $s$ being the $W$ boson mass and cms energy squared, respectively. In the numerical evaluations we take $\alpha=1/137$, $\sin^2\theta_W=0.23$, $m_W=80.38$~GeV and the $Z$ boson mass $m_Z=91.19$~GeV~\cite{Tanabashi:2018oca} throughout this article. The last term $\Delta$, explicitly given in~\ref{app_delta}, is mainly determined by the $s$-channel diagrams of~Figs.~\ref{leading}(b) and~(c). As seen in Fig.~\ref{fig_ratio}, its contribution to the overall result is relatively small, $\lesssim4\%$, and vanishes  asymptotically ($s\rightarrow\infty$). Therefore, ignoring $\Delta$ in some cases would not lead to a dramatic change in the cross section. It is crucial for our analysis that the leading contribution to the cross section is logarithmic. The logarithm comes from the diagram of Fig.~\ref{leading}(a) with a neutrino in the intermediate state~\cite{Alles:1976qv}. One can check that each term with the logarithm arises only either from the square of this diagram itself or from the interference terms with this diagram (see~\ref{app_cross}). The contributions from the rest two diagrams cancel out almost completely and one sees the presence of the corresponding processes mainly due to their providing the interference of the channel with the neutrino exchange. In the next section we will show that~$\sigma_{\text{EW}}$ is directly related to the distribution of the effective neutrinos in the electron. 

%%%%%%%%%%%%%%
% FIGURE 2
%%%%%%%%%%%%%%
%%%%%%%%%%%%%%

%%%%%%%%%%%%%%%%%%%%%%%%%%%%%%%%%%%%%%%%%%%%%%%%%%%%%%%%%
\section{The effective neutrino mechanism in $e^+e^-\rightarrow W^+W^-$\label{ena}}
%%%%%%%%%%%%%%%%%%%%%%%%%%%%%%%%%%%%%%%%%%%%%%%%%%%%%%%%
It is known that, as the quark distributions in the quark--parton model~\cite{Bjorken:1969ja}, one can introduce a distribution of effective (equivalent) electrons inside the electron after a photon emission~\cite{Chen:1975sh,kessler,Baier:1973ms}. This is schematically illustrated in Fig.~\ref{fig1p}. The probability density that the effective electron will  participate in a subprocess proceeding at the energy squared in the interval $\hat s$ to $\hat s+\mathrm{d}\hat s$ has the \newpage \noindent following form~\cite{Alikhanov:2018kpa}:

%%%%%%%%%%%%%%
% FIGURE 3
%%%%%%%%%%%%%%
\begin{figure}[b]
\centering
\includegraphics[width=0.9\textwidth]{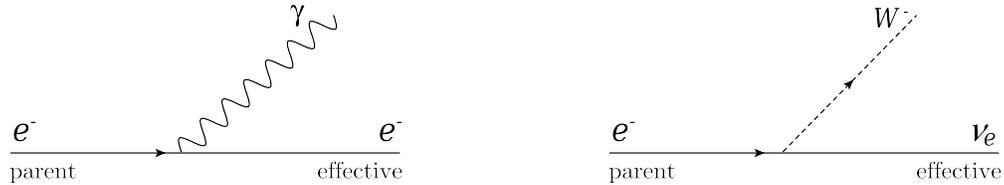}
\caption{The effective electron in the parent electron (left). The effective neutrino in the parent electron (right).}
\label{fig1p}
\end{figure}
%%%%%%%%%%%%%

\begin{equation}
f_{e/e}(\hat s,Q^2)=\frac{\alpha}{2\pi}\frac{1}{s}\left[\frac{1+y^2}{1-y}\ln\left(\frac{Q^2_{\text{max}}}{Q^2_{\text{min}}}\right)-\frac{1}{s}\left(Q^2_{\text{max}}-Q^2_{\text{min}}\right)\right],\label{eq:gen2}
\end{equation} 
where $y=\hat s/s$, $0\leq \hat s\leq s$ with $s$ being the total cms energy squared of a given reaction under study, $Q^2_{\text{min}}$ and $Q^2_{\text{max}}$ are the minimum and maximum of the four-momentum transfer squared in the reaction, respectively. A detailed discussion of such densities is given, for example, in Ref.~\citen{Chen:1975sh}. 

One can easily write the analog of~\eqref{eq:gen2} for the effective neutrinos by replacing the emitted photon by a $W^-$ boson~(see Fig.~\ref{fig1p}). The neutrino distribution in the electron will then read

\begin{equation}
f^T_{\nu/e}(\hat s,Q^2)=\frac{\alpha}{2\pi}\frac{1}{4\sin^2{\theta_W}}\frac{1}{s}\left[\frac{1+y^2}{1-y}\ln\left(\frac{Q^2_{\text{max}}+m_W^2}{Q^2_{\text{min}}+m_W^2}\right)-\frac{1}{s}\left(Q^2_{\text{max}}-Q^2_{\text{min}}\right)\right].
\label{funw}
\end{equation}
Here the $T$ superscript is to indicate that the distribution corresponds to the emission of a transversely polarized $W$ boson. Since the boson is massive, $0\leq \hat s\leq\left(\sqrt{s}-m_W\right)^2$ and the validity of the condition $\sqrt{s}>m_W$ is simultaneously assumed. In the limit $m_W\rightarrow 0$ and~$1/(4\sin^2{\theta_W})\rightarrow~1$,~\eqref{funw} reduces exactly to \eqref{eq:gen2}, as it must be. It is obvious that the distribution of the antineutrinos in the positron will also be given by~\eqref{funw} due to $CP$~invariance. Also note that \eqref{funw} transforms into the distribution of the effective $W^-$ bosons in the electron with the replacement $y\rightleftarrows (1-y)$, being consistent in the high energy limit with the spectrum from Refs.~\citen{Kane:1984bb, Dawson:1984gx}.  This is similar to the property that the Weizs\"acker--Williams equivalent photon spectrum of the electron can be obtained from~\eqref{eq:gen2} by simply exchanging the electron and photon momenta~\cite{Chen:1975sh}.

The Glashow resonance in $e^+e^-\rightarrow W^+W^-$ is excited when the incident electron annihilates with an effective antineutrino from the positron, as displayed in~Fig.~\ref{fig2p}. The corresponding cross section can then be represented as a convolution over~$\hat s$: 

%%%%%%%%%%%%%%
% FIGURE 4
%%%%%%%%%%%%%%
\begin{figure}[b]
\centering
\includegraphics[width=0.8\textwidth]{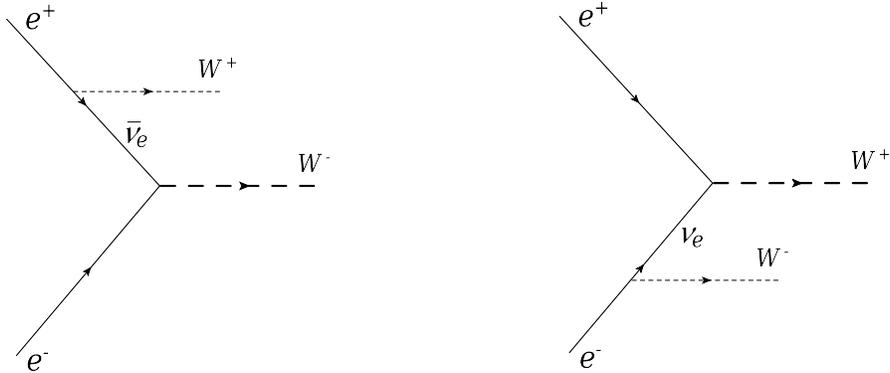}
\caption{The Glashow resonance (left) and its $CP$ conjugate (right) in $e^+e^-\rightarrow W^+W^-$. The resonance is excited when the incident electron annihilates with an effective antineutrino from the positron.}
\label{fig2p}
\end{figure}
%%%%%%%%%%%%%

\begin{equation}
\sigma=2\int_{0}^{\left(\sqrt{s}-m_W\right)^2}f^T_{\nu/e}(\hat s,Q^2)\sigma_{\nu e\rightarrow W}(\hat s)\mathrm{d}\hat s,\label{main_sec}
\end{equation}
where $\sigma_{\nu e\rightarrow W}(\hat s)$ is the cross section for the contributing subprocesses $\bar\nu_ee^-\rightarrow W^-$ and $\nu_ee^+\rightarrow W^+$. The factor 2 implies that the distributions for $\nu_e$ in $e^-$ and $\bar\nu_e$ in $e^+$ are equal to each other. In order to evaluate the integral in~\eqref{main_sec} we adopt the narrow width approximation: 

\begin{equation}
\sigma_{\nu e\rightarrow W}(\hat s)=\frac{2\pi^2\alpha}{\sin^2{\theta_W}}\delta(\hat s-m_W^2), \label{narrow}
\end{equation}
where $\delta(x)$ is the Dirac delta function.
Substituting~\eqref{funw} and~\eqref{narrow} into~\eqref{main_sec} we obtain

\begin{equation}
\sigma=\frac{\pi\alpha^2}{2\sin^4{\theta_W}}\frac{1}{s}\left[\frac{1+\tau^2}{1-\tau}\ln\left(\frac{1+\beta}{1-\beta}\right)-\beta\right],
\label{funena}
\end{equation} 
where we have also used that $Q^2_{\genfrac{}{}{0pt}{}{\text{max}}{\text{min}}}=s(1\pm\beta)/2-m_W^2$, which follows simply from kinematics of the process $e^+e^-\rightarrow W^+W^-$. Note that the cross section~\eqref{funena} is a projection of~\eqref{funw}, i.e.~$\sigma\propto f^T_{\nu/e}(\tau,Q^2)$.

%%%%%%%%%%%%%%
% FIGURE 5
%%%%%%%%%%%%%%
\begin{figure}[b]
\centering
\includegraphics[width=0.7\textwidth]{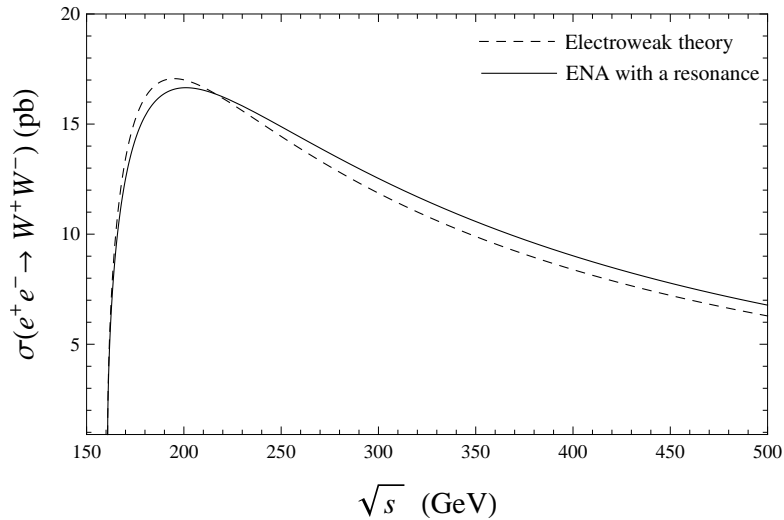}
\caption{The total cross section for~$e^+e^-\rightarrow W^+W^-$ as a function of the cms energy. The dashed curve is the exact result obtained in the electroweak theory, Eq.~\eqref{funfel}. The solid curve is the prediction of the effective neutrino approach with the Glashow resonance, Eq.~\eqref{relation}.} 
\label{fig3p}
\end{figure}
%%%%%%%%%%%%%% 
Returning to the electroweak result one can see that~\eqref{funena} reproduces the analytical structure of~\eqref{funfel} with minor~deviations. It is therefore natural to suppose that the electroweak theory reflects the presence of the resonant mechanism in~$e^+e^-\rightarrow W^+W^-$.  In other words, $\sigma_{\text{EW}}$ involves in~fact the projection of the neutrino distribution distorted slightly by the contribution from the diagrams of~Figs.~\ref{leading}(b) and~(c). If one ignores the distortion along with the term~$\Delta$ (whose contribution, as we have shown above, is a few percent), then $\sigma_{\text{EW}}=\sigma$ and~\eqref{funfel} takes a compact and interpretable form:

\begin{equation}
\sigma_{\text{EW}}=\pi g^2 f^T_{\nu/e}(\tau,Q^2),\label{relation}
\end{equation}
where $g=\sqrt{4\pi\alpha}/\sin{\theta_W}$ is the electroweak gauge coupling. In Fig.~\ref{fig3p} we demonstrate numerically that~\eqref{relation} is actually in good agreement with~\eqref{funfel}. Furthermore, the agreement becomes exact asymptotically~\cite{Alikhanov:2018xdi}. The ENA provides thus an interpretation of the nature of the cross section for~$e^+e^-\rightarrow~W^+W^-$: it is the distribution of the effective neutrinos in the electron mapped by the Glashow resonance. Hence the process~$e^+e^-\rightarrow~W^+W^-$ may be envisaged as proceeding through the Glashow resonance with a relatively small admixture of the $s$-channel $\gamma$ and $Z$ exchanges. 

It is interesting to note that a similar situation has been known to occur with the excitation of narrow neutral resonances in electron--positron collisions accompanied by initial state radiation (ISR)~\cite{Baier:1969kaa}. In this case, though a resonance appears, the cross section also does not have a distinct Breit--Wigner peak but rather behaves as a function falling smoothly with  energy.~At~the~same~time, the latter function turns out to be a projection of the distribution of the effective (equivalent) electrons in the electron,~$\sigma_{\text{ISR}}\propto f_{e/e}$~\cite{Chen:1975sh}.

%%%%%%%%%%%%%%%%%%%%%%%%%%%%%%%%%%%%%%%%%%%%%%%%
\section{Comparison with experiment\label{experiment}}
\begin{figure}
\centering
\includegraphics[width=0.7\textwidth]{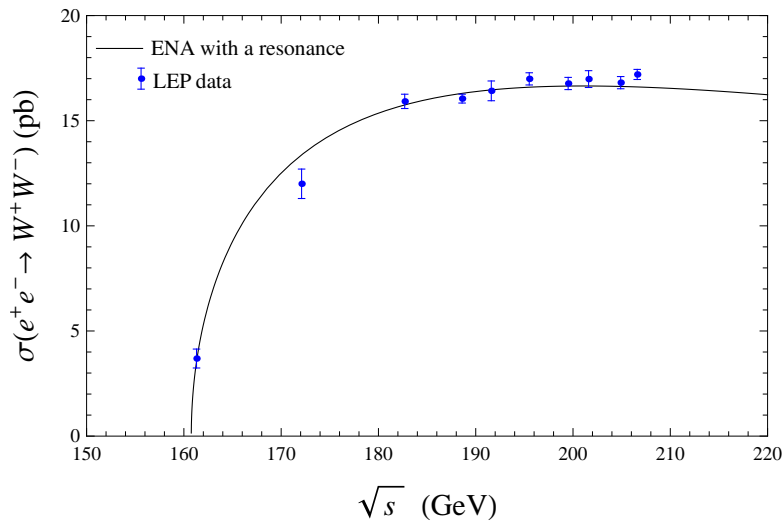}
\caption{The total cross section for~$e^+e^-\rightarrow W^+W^-$ as a function of the cms energy.  The solid curve is the prediction of the effective neutrino approach with the Glashow resonance, Eq.~\eqref{relation}. The points are the combined results of the LEP experiments ALEPH, DELPHI, L3 and OPAL~\cite{Schael:2013ita}.}
\label{fig4p}
\end{figure}

%%%%%%%%%%%%%%%%%%%%%%%%%%%%%%%%%%%%%%%%%%%%%%
The cross section for $e^+e^-\rightarrow W^+W^-$ was measured at the electron--positron collider~LEP at CERN by the four experiments ALEPH, DELPHI, L3 and OPAL in the energy region from slightly above the threshold to $\sqrt{s}=206.6$~GeV~\cite{Schael:2013ita}. In Fig.~\ref{fig4p} we compare the prediction~\eqref{relation} with the results from LEP. The effective neutrino mechanism gives a good fit to the experimental data and allows one to interpret the LEP measurements as an observation of the Glashow resonance. 

%%%%%%%%%%%%%%
% FIGURE 6
%%%%%%%%%%%%%%
%%%%%%%%%%%%%%

%%%%%%%%%%%%%%%%%%%%%
\section{Conclusions \label{concl}}
%%%%%%%%%%%%%%%%%%%%
A rapid enhancement of the $\bar\nu_ee^-$ scattering cross section at the $W^-$ boson production threshold, commonly referred to as the Glashow resonance, is a prediction of the Standard Model that remains so far unidentified experimentally. The IceCube neutrino telescope has detected events that could in principle be initiated by the resonance, however they do not allow to make a decisive conclusion. 

Recently it was pointed out that the process~$e^+e^-\rightarrow~W^+W^-$ may be envisaged as proceeding through the Glashow resonance and its $CP$ conjugate~\cite{Alikhanov:2018xdi}.
In this article we have investigated a possible contribution of the resonance to~$e^+e^-\rightarrow~W^+W^-$ in the whole energy range starting from the threshold~$\sqrt{s}=2m_W$. By extending the method of the effective (equivalent) particles to the neutral leptons, we relate the distribution of the effective neutrinos in the electron to the total cross section for~$e^+e^-\rightarrow~W^+W^-$. We show that the cross section is the neutrino distribution mapped by the Glashow resonance. The mapping is distorted by a small contribution from the other channels not involving neutrinos.  Our approach gives a good fit to the existing experimental data measured at LEP and allows one to interpret these measurements as an observation of the Glashow resonance. 

It should be emphasized that in the high energy limit the exact electroweak result for the cross section coincides with that given by the effective neutrino mechanism. The latter means that the admixture of the leading non-resonant channels in $e^+e^-\rightarrow W^+W^-$ decreases with increasing cms energy. Already at $\sqrt{s}=500$~GeV this is $\lesssim 2\%$ and drops down below $1\%$ at $\sqrt{s}=3$~TeV. Such energies can be achieved at future electron--positron colliders  as the International Linear Collider (ILC)~\cite{Fujii:2015jha} and the Compact Linear Collider (CLIC)~\cite{CLIC:2016zwp}, where the Glashow resonance could therefore manifest itself more purely in the $W^+W^-$ pair final states.  

%%%%%%%%%%%%%%%%%%%%%%%%%%
\section*{Acknowledgments}
%%%%%%%%%%%%%%%%%%%%%%%%%%
\noindent I would like to thank Professor E. A.~Paschos for useful comments and discussions.
This work was partly supported by the Program of fundamental scientific research of the Presidium of the Russian Academy of Sciences "Physics of fundamental interactions and nuclear technologies".

%%%%%%%%%%%%%%%%%%%%%%%%%
\appendix\section{\label{app_delta}}
%%%%%%%%%%%%%%%%%%%%%%%%%

\noindent The explicit form of the term $\Delta$ in~\eqref{funfel} is
\begin{eqnarray}
\Delta=\frac{\pi\alpha^2}{2s_W^4}\frac{\beta}{s}\left\{\frac{m_Z^2(1-2s_W^2)}{s-m_Z^2}\left[(4\tau+2\tau^2)\frac{L}{\beta}-\frac{1+20\tau+12\tau^2}{12\tau}\right]\right.\nonumber\\+\left.\frac{\beta^2m_Z^4(8s_W^4-4s_W^2+1)(1+20\tau+12\tau^2)}{48\tau^2(s-m_Z^2)^2}\right\}
\end{eqnarray}
with $s_W\equiv\sin\theta_W$, $L\equiv\ln\left[(1+\beta)/(1-\beta)\right]$.

%%%%%%%%%%%%%%%%%%%%%%%%%%
\section{\label{app_cross}}
%%%%%%%%%%%%%%%%%%%%%%%%%

\noindent The total cross section for $e^+e^-\rightarrow W^+W^-$ calculated within the electroweak theory is~\cite{Alles:1976qv}

\begin{eqnarray}
\sigma_{\text{EW}}=\frac{\pi\alpha^2}{8s_W^4}\frac{\beta}{s}\left[\sigma_{\nu\nu}+\sigma_{\gamma\gamma}+\sigma_{ZZ}+\sigma_{\nu\gamma}+\sigma_{\nu Z}+\sigma_{\gamma Z}\right],
\end{eqnarray}
where

\begin{eqnarray}
\sigma_{\nu\nu}=\sigma_1,\,\,\,\,\,\,\,\,\,\,\sigma_{\gamma\gamma}=s_W^4\sigma_2,\,\,\,\,\,\,\,\,\,\,
\sigma_{ZZ}=\left(s_W^4-\frac{1}{2}s_W^2+\frac{1}{8}\right)\frac{s^2}{(s-m_Z^2)^2}\sigma_2,
\end{eqnarray}

\begin{eqnarray}
\sigma_{\nu\gamma}=-s_W^2\sigma_3,\,\,\,\,\,\,\,\,\,\,\sigma_{\nu Z}=\left(s_W^2-\frac{1}{2}\right)\frac{s}{(s-m_Z^2)}\sigma_3,
\end{eqnarray}

\begin{eqnarray}
\sigma_{\gamma Z}=-2s_W^2\left(s_W^2-\frac{1}{4}\right)\frac{s}{(s-m_Z^2)}\sigma_2
\end{eqnarray}
with

\begin{eqnarray}
\sigma_{1}=\frac{2}{\tau}+\frac{\beta^2}{12\tau^2}+4\left[\left(1-2\tau\right)\frac{L}{\beta}-1\right],
\end{eqnarray}

\begin{eqnarray}
\sigma_{2}=\frac{16\beta^2}{\tau}+\frac{2\beta^2}{3}\left[\frac{1}{\tau^2}-\frac{4}{\tau}+12\right],
\end{eqnarray}

\begin{eqnarray}
\sigma_{3}=16-32\tau\frac{L}{\beta}+\frac{8\beta^2}{\tau}+\frac{\beta^2}{3\tau^2}\left(1-2\tau\right)+4\left(1-2\tau\right)-16\tau^2\frac{L}{\beta}.
\end{eqnarray}
Here, for example, $\sigma_{\nu\nu}$ and $\sigma_{\gamma\gamma}$ correspond to the contributions from the squares of the diagrams with the neutrino and the photon~of Figs.~\ref{leading}(a) and~(b), respectively, $\sigma_{\nu\gamma}$ comes from the interference between these diagrams, and so on. Adding up all the contributions yields~\eqref{funfel}.

%\begin{thebibliography}{000} %for 3 digits
%\begin{thebibliography}{00}  %for 2 digits

\end{document}